\documentstyle[11pt,newpasp,twoside,epsf]{article}
\def\edcomment#1{\iffalse\marginpar{\raggedright\sl#1\/}\else\relax\fi}
\reversemarginpar

\begin{document}
\title{Abundance Patterns in Planetary Nebulae}
 
\author{Karen B. Kwitter}
 
\affil{Williams College, Astronomy Department, 33 Lab
 Campus Dr., Williamstown MA 01267} 

\author{Richard B.C. Henry} 

\affil{Department of Physics \& Astronomy, University of Oklahoma,
Nielson Hall, Norman, OK 73019}

\begin{abstract}
We previously determined abundances of He, C, N, O, and Ne for a sample
of planetary nebulae (PNe) representing a broad range in progenitor
mass and metallicity, and we now compare them with theoretical predictions of
PNe abundances from a grid of intermediate-mass star models. We find
very good agreement between observations and theory, lending strong
support to our current understanding of nucleosynthesis in stars below
8M$_{\odot}$ in birth mass. In particular, C and N abundance patterns are
consistent with the occurrence of hot-bottom burning in stars \newline above
roughly 3.5$M_{\odot}$, a process that converts much of $^{12}$C into
$^{14}$N during the AGB phase. This agreement also supports the
validity of published stellar yields of C and N in the study of the
abundance evolution of these elements.

\end{abstract}

\section{Introduction}

Massive stars ($>$8M$_{\odot}$) are the principal, and in most cases,
the sole, source of elements beyond He. However, for the elements C
and N, origins are more ambiguous. Intermediate-mass stars (IMS;
1$\le$M$\le$8~M$_{\odot}$) are hot enough in their cores and fusion
shells to produce C via He-burning and N via the CNO cycle. Recent
theoretical results indicate significant C and N production in IMS
(van den Hoek \& Groenewegen 1997 (VG); Marigo, Bressan, \& Chiosi
1996 (MBC), 1998). Likewise, massive stars, too, synthesize and expel
significant C and N (Woosley \& Weaver 1995; Nomoto et al. 1997;
Maeder 1992). Over the whole stellar mass range, the general
conclusion is that N comes predominantly from IMS, while both IMS and
massive stars contribute to C.

We compare the set of abundances we determined for a sample of 20 PNe
over a broad range in progenitor mass and metallicity with PN
abundances predicted from stellar yield calculations of VG and MBC.

\section{Abundance Calculations}

The heart of our method for determining abundances is the standard one
in which abundances of observable ions for an element are first
determined using a 5-level atom calculation for each ion. Then these
ionic abundances are summed together and multiplied by an ionization
correction factor (ICF) which adjusts the sum upward to account for
unobservable ions. Finally, this product is in turn multiplied by a
model-determined factor $\xi$ which acts as a final correction to our
elemental abundance. Our modelling method has been discussed in detail
most recently in Henry, Kwitter, \& Dufour (1999). Our abundance
results along with nebular diagnostics are contained in Henry,
Kwitter, \& Howard (1996), Kwitter \& Henry (1996, 1998) and Henry \&
Kwitter (1999) Results for the entire sample are reported in full in
Table 6 of Henry \& Kwitter (1999).

\section{IMS Nucleosynthesis: Models Versus Observations}

Compilations of observed abundances in PNe, such as those by Henry
(1990) and Perinotto (1991) provide strong evidence that IMS
synthesize He, C, and N. We can infer that directly by comparing
abundance patterns in our PN sample with patterns in the interstellar
medium, i.e. H~II regions and stars.  The two figures show log(C/O)
and log(N/O) vs. 12+log(O/H), respectively, for our PN sample (filled
diamonds) along with Galactic and extragalactic H~II region data (open
circles) compiled and described in Henry \& Worthey (1999) and F and G
star data (open triangles; left-hand figure only) from Gustafsson et
al. (1999). Also shown are the positions for the sun (S; Grevesse et
al. 1996), Orion (O; Esteban et al. 1998), and M8 (M; Peimbert et
al. 1993; lef-hand figure only).

Note in the left-hand figure that in contrast to the relatively close
correlation between C and O displayed by the H~II regions and stars,
there is no such relation indicated for PNe.  In fact the range in C
is over 2.5 orders of magnitude, far greater than for the H~II
regions and stars and larger than can be explained by uncertainties
in the abundance determinations. In addition, C levels in PNe appear
on average higher than those typical of H~II regions for the same O
value, indicating that additional C above the general interstellar
level present at the time these stars formed, was produced during their
lifetimes. The right-hand figure shows similar behavior for N:
H~II regions seem to suggest a relation between N/O and O/H in the
interstellar medium, yet we see no such pattern for PNe. Also, N/O tends
to be systematically higher for PNe than for H~II regions, again
suggesting that N is produced by PN progenitors.

These figures imply that C and N are synthesized in IMS; evidence from
Ne/O strengthens this contention. Limited space prohibits inclusion of
a similar figure that shows a constant value for Ne/O over a range in
O abundance; the pattern displayed by PNe is indistinguishable from
that of the H~II regions. For a full discussion see Henry \& Worthey
(1999).

\begin{figure}
\plotfiddle{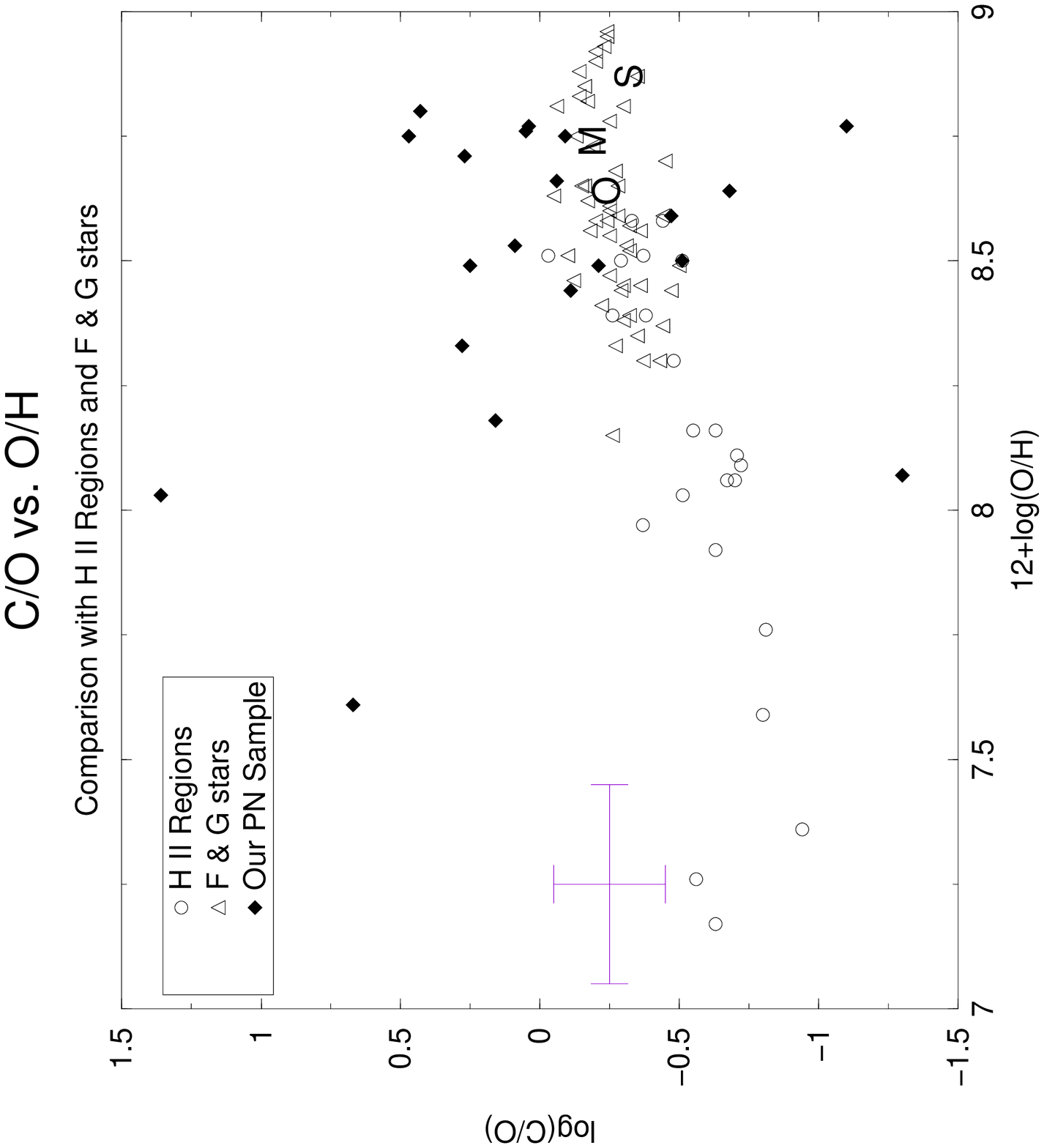}{1.1in}{270}{35}{40}{-220}{120}
\plotfiddle{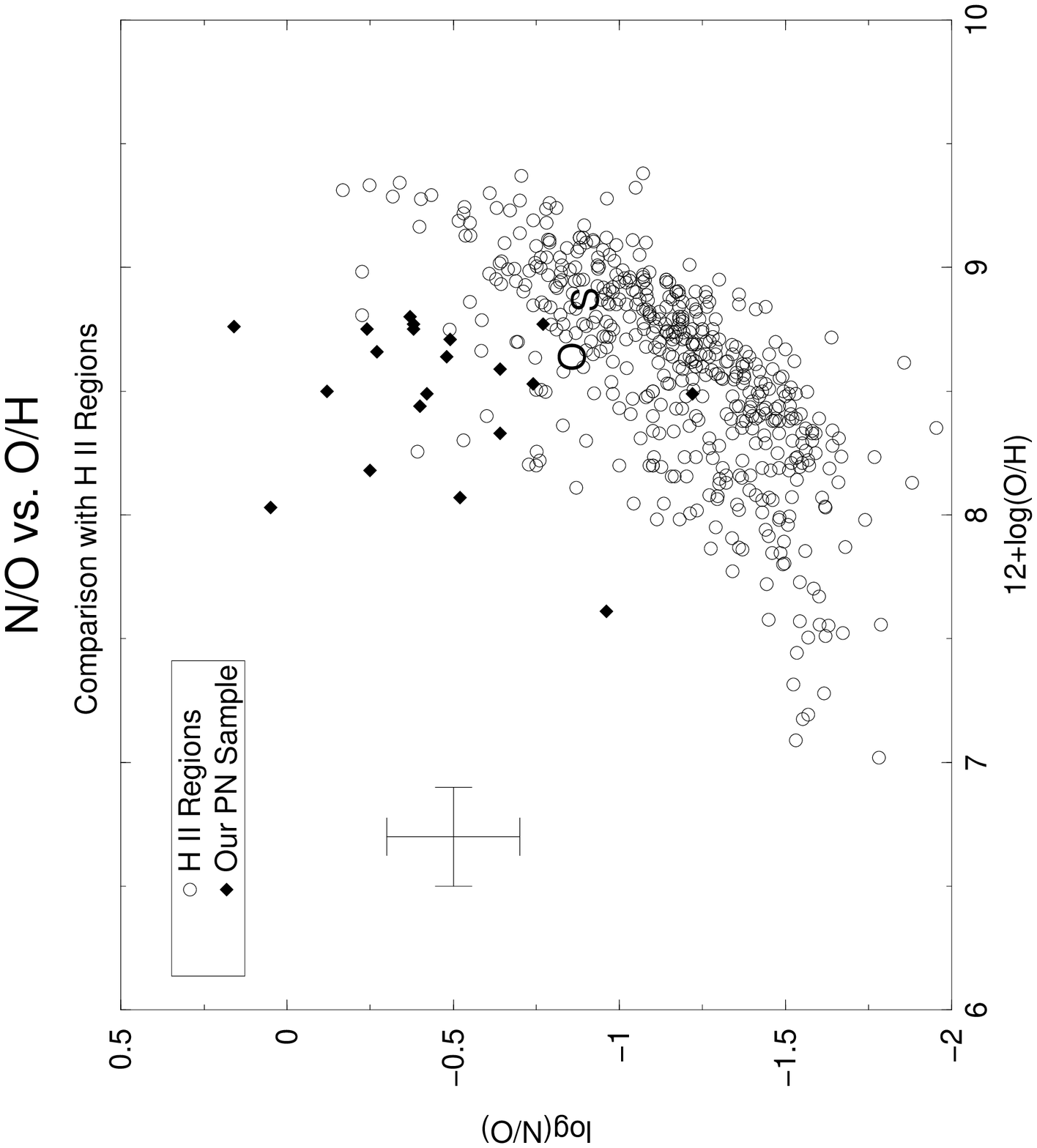}{1.1in}{270}{35}{40}{0}{210}
\end{figure}

\section{Comparison with Predicted Yields}

We used our PN abundance results to test the theoretical predictions
of PN abundances; for details see Henry \& Kwitter (1999). We tested
two published sets of theoretical calculations. VG calculated a grid
of stellar models ranging in mass fraction metallicity between 0.001
and 0.04 and progenitor mass of 0.8 to 8~M$_{\odot}$. Likewise, MBC
calculated models for mass fraction metallicity of 0.008 and 0.02 for
stars between 0.7 and 5~M$_{\odot}$. Both teams employed up-to-date
information about opacities and mass loss to calculate yields for
several isotopes, including $^{4}$He, $^{12}$C, $^{13}$C, and
$^{14}$N.

The progenitor metallicity range consistent with our results is
between 1/20 solar and solar. We found that observed abundances of C
vs. O are consistent with predictions for both high- and low- mass
progenitors. At all metallicities the C abundance is initially
predicted to rise with mass but then drop back to low values as mass
continues to increase above 2-3~M$_{\odot}$. This reversal is the
result of hot-bottom burning in stars with greater masses than this in
which C from the 3rd dredge-up is converted to N at the
base of the convective envelope late in the AGB stage.

For N vs. O, the predicted behavior with progenitor mass is positively
monotonic and is consistent with our abundances. Apparently the C and
N abundances observed in PNe are consistent with progenitor masses in
the range of 1-4~M$_{\odot}$.

Consideration of N vs. He reinforces our conclusions about the PN
progenitor mass range. Theoretical abundance predictions for
progenitors in the 1-4~M$_{\odot}$ range are consistent with our
observations. The one extreme outlier is PB6, whose unusually high
He abundance (He/H=0.20) needs to be confirmed independently.

Our detailed comparison of observed PN abundances with predicted ones
has demonstrated good agreement between the two and is indeed
encouraging. To the extent to which predicted PN abundances are
related in turn to the actual stellar yields, our comparison provides
what we believe to be the best empirical support yet for the
theoretical calculations. It is imperative, however, that these models
continue to be tested with larger samples of PNe whose abundances have
been carefully determined. As this is done, we will be better able to
ascertain the exact role that intermediate mass stars play in the
synthesis of C, N, and He in galaxies.    

\section{Summary}

\begin{itemize}
 
\item Abundances of C and N in PNe, when plotted against
O, show a much broader range than H~II regions and F and G stars
and are generally higher. At the same time, both O and Ne
display similar patterns in both PNe and H~II regions. Taken together,
these results support the idea the PN progenitors synthesize
significant amounts of C and N.
 
\item Abundances of C, N and He found in our sample of PNe are
consistent with model predictions. We believe that this is the first
time that such a detailed comparison of observation and theory has
been possible and that the results provide encouragement for the use
of published yields of intermediate mass stars in studying galactic
chemical evolution, especially in the cases of C and N.
 
\item Our comparisons of observed and predicted PN abundances support the occurrence of hot-bottom burning in stars above about 3.5-4~M$_{\odot}$.
 
\end{itemize}

\acknowledgments
We are grateful to the support staff at KPNO for help in carrying out
the observing portions of this program.  This project was supported by
NASA grant NAG 5-2389.


\begin{references}

\reference Esteban, C., Peimbert, M., Torres-Peimbert, S., \& Escalante, V. 1998, \mnras, 295, 401
\reference  Grevesse, N., Noels, A., \& Sauval, A.J. 1996, in ASP Conf. Ser. 99, Cosmic Abundances, ed. S.S. Holt \& G. Sonneborn (San Francisco: ASP), 117
\reference Gustafsson, B., Karlsson, T., Olsson, E.,
Edvardsson, B., \& Ryde, N. 1999, \aap, 342, 426
\reference Henry, R.B.C. 1990, \apj, 356, 229
\reference Henry, R.B.C., Kwitter, K.B., \& Dufour, R.J. 1999, \apj, 517, 782
\reference Henry, R.B.C., Kwitter, K.B., \& Howard, J.W. 1996 \apj, 458, 215
\reference Henry, R.B.C., \& Worthey, G. 1999, \pasp, 111, 919
\reference van den Hoek, L.B., \& Groenewegen, M.A.T. 1997, \aaps, 123, 305
\reference Henry, R.B.C., \& Kwitter, K.B. 1999, \apj, submitted
\reference Kwitter, K.B., \& Henry, R.B.C. 1996, \apj, 473, 304
\reference Kwitter, K.B., \& Henry, R.B.C. 1998, \apj, 493, 247
\reference Maeder, A. 1992, \aap, 264, 105
\reference Marigo, P., Bressan, A., \& Chiosi, C. 1996, \aap, 313, 545
\reference Marigo, P., Bressan, A., \& Chiosi, C. 1998, \aap, 331, 564
\reference Nomoto, K., Hashimoto, M., Tsujimoto, T., Thielemann, F.-K., Kishimoto, N., Kubo, Y., \& Nakasato, N. 1997, Nuc. Phys. A, A616, 79c
\reference Peimbert, M., Torres-Peimbert, S., \& Dufour, R.J. 1993, \apj, 418, 760
\reference Perinotto, M. 1991, \apjs, 76, 687
\reference Woosley, S.E., \& Weaver, T.A. 1995, \apjs, 101, 181


\end{references}
\end{document}